\documentclass[twocolumn]{article}  
\usepackage{graphicx}
\usepackage[T1]{fontenc}
\usepackage[subrefformat=parens]{subcaption}

\makeatletter
\let\cl@chapter\undefined
\makeatletter

\usepackage{amsmath,amssymb}   
\usepackage{amsthm} 
\usepackage{mathtools} 
\usepackage[numbers,sort&compress]{natbib}
\usepackage[margin=20mm]{geometry}
\usepackage{hyperref}
\hypersetup{colorlinks=true,pdfborder={0 0 0}}
\usepackage{cleveref}  
\usepackage{amsbsy}
\usepackage{bm}
\usepackage{url}
\usepackage{color}
\usepackage{multirow}
\usepackage{booktabs}
\usepackage{autobreak}
\allowdisplaybreaks 

\newcommand{\inlineTag}{
    \refstepcounter{equation}
    \bgroup\normalfont\normalcolor (\theequation)\egroup}

\crefname{equation}{Eq.}{Eqs.}%
\crefname{figure}{Fig.}{Figs.}%

\usepackage{algorithm}
\usepackage[noend]{algpseudocode}

\algnewcommand{\Break}{\textbf{break}}
\algrenewcommand\algorithmicindent{1em}

\providecommand{\argmin}{\operatornamewithlimits{argmin}} 

\everydisplay{\small} 

\newcommand\Erase{\bgroup\markoverwith{\textcolor{red}{\rule[.5ex]{2pt}{0.4pt}}}\ULon}

\begin{document}

    \title{
        Data Augmentation Methods of Dynamic Model Identification for Harbor Maneuvers using Feedforward Neural Network
    }
    
    \author{
        Kouki Wakita$^{1}$ \and
        Yoshiki Miyauchi$^{1}$ \and
        Youhei Akimoto$^{2,3}$ \and
        Atsuo Maki$^{1}$
    }

    \date{%
    \flushleft{\footnotesize
        $^1$Osaka University, 2-1 Yamadaoka, Suita, Osaka, Japan \\%
        $^2$Faculty of Engineering, Information and Systems, University of Tsukuba, 1-1-1 Tennodai, Tsukuba, Ibaraki 305-8573, Japan \\
        $^3$RIKEN Center for Advanced Intelligence Project, 1-4-1 Nihonbashi, Chuo-ku, Tokyo 103-0027, Japan \\[2ex]%
        Keywords:System Identification; Recurrent Neural Network; Automatic Berthing; Random Maneuver\\[1ex]%
        Email: kouki\_wakita@naoe.eng.osaka-u.ac.jp; maki@naoe.eng.osaka-u.ac.jp \\
    }}


    \maketitle

    \begin{abstract}
        A dynamic model for an automatic berthing and unberthing controller has to estimate harbor maneuvers, which include berthing, unberthing, approach maneuvers to berths, and entering and leaving the port. When the dynamic model is estimated by the system identification, a large number of tests or trials are required to measure the various motions of harbor maneuvers. However, the amount of data that can be obtained is limited due to the high costs and time-consuming nature of full-scale ship trials. In this paper, we improve the generalization performance of the dynamic model for the automatic berthing and unberthing controller by introducing data augmentation. This study used slicing and jittering as data augmentation methods and confirmed their effectiveness by numerical experiments using the free-running model tests. The dynamic model is represented by a neural network-based model in numerical experiments. Results of numerical experiments demonstrated that slicing and jittering are effective data augmentation methods but could not improve generalization performance for extrapolation states of the original dataset.
    \end{abstract}
    
    \section{Introduction}\label{sec:intro}
        A dynamic maneuvering model for a ship is essential for the research and development of maritime autonomous surface ships (MASS). Dynamic models can predict and imitate ship maneuvering motions in numerical simulations. Therefore, the dynamic model enables the design or tuning of control algorithms and the validation of automatic navigation systems.

        One of the technical challenges in achieving MASS is the development of the automatic berthing and unberthing controller, which requires an appropriate dynamic model. To achieve berthing, the ship's speed needs to be significantly reduced before reaching the berth. In addition, various maneuvering motions such as turning tightly, astern, and crabbing may be required depending on the port and shape of the berth. In essence, a dynamic model for the automatic berthing and unberthing controller is necessary to estimate harbor maneuvers, which include various maneuvers like berthing and unberthing, approach maneuvers to berths, and entering and leaving the port.
        Moreover, when ships are controlled by algorithms like optimal control or reinforcement learning, maneuvers that do not exist in human operations may occur, and the dynamic model has to evaluate such maneuvers properly. Therefore, a dynamic model for the automatic berthing and unberthing controller is desirable to estimate all possible ship motions at low speeds.

        Numerous studies have been conducted on the dynamic model for the ship maneuvering motion. In particular, many studies have focused on the modeling of the hydrodynamic force acting on the ship.
        For instance, one of the dynamic models for the ship maneuvering motion is the Abkowitz model \cite{Abkowitz1980}. The Abkowitz model represents the hydrodynamic force by polynomials. These polynomials are obtained from a Taylor expansion about the forward motion condition. The model is simple to derive and can be added with nonlinear terms, which enables it to represent a wider range of ship maneuvering behaviors.
        Besides, the MMG model, which was proposed by a research group of the Japan Towing Tank Conference \cite{Ogawa1978}, is also one of the dynamic models applicable for simulating ship motions. The MMG model is a modular-type mathematical model consisting of submodels, which express the hydrodynamic force induced by the hull, propeller, rudder, and other actuators. Thus, the MMG model only requires modification of the relevant submodels even if partial design changes occur, such as changes to the rudder. 

        The standard MMG model \cite{Yasukawa2015} and classical Abkowitz model \cite{Abkowitz1980} are primarily focused on ship maneuvering in which the ship's forward speed is sufficiently large and steady. These models are not supposed to predict harbor maneuvers. However, several studies have been conducted on the modification of the MMG model for low-speed maneuvering \cite{Kose1984,Fujino1990,KOBAYASHI1994,ISHIBASHI1996,Yoshimura2009}. Miyauchi et al. expanded the Abkowitz model for harbor maneuver \cite{Miyauchi2021JAS} and then proposed an automatic derivation method of a hybrid model combining the MMG and Abkowitz models for harbor maneuvering motions \cite{Miyauchi2023}. Those studies have enabled the MMG and Abkowitz models to represent a greater variety of maneuvering behaviors. 

        Other studies have used neural networks (NNs) \cite{MOREIRA2003,Chiu2004,RAJESH2008,OSKIN2013,Wakita2022} and kernel functions \cite{Zhang2011,LUO2014} for estimating the dynamic model for the ship's maneuver although these studies did not necessarily focus on harbor maneuvers. These models do not have a physical and hydrodynamic background but have certain advantages in their high approximation capabilities \cite{Cybenko1989,HORNIK1991251}.

        The dynamic models have many parameters, that is, hydrodynamic coefficients, that have to be determined according to the ship. Captive model tests \cite{Yasukawa2015} and empirical formulas \cite{SUKAS2019} are often used to determine the parameters. However, captive model tests require special test facilities and a significant amount of time and effort for experimentation, depending on the complexity of the model. Therefore, system identification (SI) is also often used as an alternative method. 
        SI techniques allow parameter identification from the full-scale ship trial or free-running model tests. In other words, SI techniques require time series data of state and control variables to identify parameters; thus, force/moment measurements by captive model tests are not required in principle.
        Many studies have been conducted on the SI of dynamic models for the ship's maneuver, and various identification methods have been applied \cite{Abkowitz1980,ASTROM1976,YOON2003,ARAKI2012,SUTULO2014,Miyauchi2022,Wakita2022}.
        {\AA}str\"{o}m and K\"{a}llstr\"{o}m \cite{ASTROM1976,KALLSTROM1981} applied the maximum likelihood method to determine the ship steering dynamics from measurements of a freighter and a tanker. 
        Abkowitz \cite{Abkowitz1980} proposed the identification method for the hydrodynamic coefficients and tidal current state based on the extended Kalman filter.
        Yoon and Rhee \cite{YOON2003} used ridge regression to estimate the hydrodynamic coefficient of the polynomial model from sea trial data smoothed by modified Bryson-Frazier smoother. 
        Sutulo et al. \cite{SUTULO2014} proposed an identification method based on the classic genetic algorithm used to minimize the distance between the observed state history and time series data recovered by maneuvering simulation.
        Miyauch et al. \cite{Miyauchi2022} applied the covariance matrix adaptation evolution strategy to explore the system parameters of the MMG model using the free-running model tests.
        
        SI requires an appropriate dataset in accordance with the purpose and complexity of the dynamic model. For instance, the parameter identification using zigzag and turning tests may suffer from multicollinearity due to the strong correlation between the sway and yaw angle velocity \cite{Abkowitz1980}. Several studies have been conducted on the optimal design of the control input for measuring the training data that improves the stability and accuracy of parameter identification of the dynamic model \cite{YOON2003,WANG2020}. 
        
        Most SI studies for a ship's maneuvering model use zigzag and turning maneuvers of the full-scale trials or free-running model tests as training datasets. However, zigzag and turning tests are not suitable for estimating the dynamic model for harbor maneuvers since they cannot measure harbor maneuvers, such as astern and crabbing motions. Therefore, several previous studies \cite{Miyauchi2022,Wakita2022,Miyauchi2023} used the dataset measured in random maneuvers. Note that random maneuvers are maneuvers in which control inputs are manually selected to obtain the various values of state variables and control inputs, including harbor maneuvers.
        
        Identifying a dynamic model for the automatic berthing and unberthing controller requires conducting a large number of tests or trials and measuring various motions. However, conducting numerous full-scale ship trials is impractical due to the high cost involved. Therefore, a method for identifying a dynamic model with a limited amount of data is required. One of the options to achieve the high generalization performance of dynamic models is to apply data augmentation, which generates synthetic data.

        Many studies have been conducted on data augmentation methods that can be applied to time series data \cite{Wen2021,Iwana2021}. However, to the best of our knowledge, no studies have been conducted on applying data augmentation methods to the parameter identification of dynamic models representing ship maneuvering motions. In addition, all data augmentation methods do not necessarily improve the generalization performance of the ship's dynamic model. In fact, many generation methods of synthetic data distort the meaning of the original data. For example, window wrapping \cite{leguennec2016,Um2017,IsmailFawaz2018}, which is one of the time wrapping methods, generates synthetic data by compressing or stretching time series data. This should not be applied to this problem since the time derivative of the state variable predicted by the dynamic model may be changed significantly. These data augmentation methods may cause the deterioration of the generalization performance by generating data whose characteristics are different from that of the original data.

        This study aims to improve the generalization performance of the dynamic model for an automatic berthing and unberthing controller by introducing data augmentation. For this purpose, this study demonstrates effective data augmentation methods and their effectiveness.
        In this paper, slicing \cite{Cui2016,leguennec2016} and jittering \cite{Bishop1995,Um2017,RASHID2019} are introduced as the data augmentation methods. The application of these data augmentation methods to the parameter identification problem of dynamic models is described. Numerical experiments are conducted to demonstrate the effectiveness of the data augmentation methods. In this study, the dynamic model is represented by the NN-based model, and the datasets were measured in free-running model tests of the random maneuvers, the same as the previous studies \cite{Miyauchi2022,Wakita2022,Miyauchi2023}.
        The contents of this paper overlap the previous literature \cite{Wakita2023_conference}, but presents the results more extensively, with some revisions.
        
        The remainder of this paper is organized as follows: \Cref{sec:problem} defines the parameter identification methods of the dynamic model; \Cref{sec:aug} describes data augmentation methods; \Cref{sec:exp} shows numerical experiments to show the effectiveness of the data augmentation methods; \Cref{sec:discuss} discuss the numerical experiments and future works; finally, \Cref{sec:conclude} concludes this paper.
        
    \section{Parameter Identification Methods}\label{sec:problem}
        In this section, we describe the parameter identification methods for the dynamic model using time series data obtained from full-scale ship trials or free-running model tests. In this study, the metric of the prediction error of the dynamic model was defined, and the parameter identification method was formulated as a minimization problem of the prediction error. The parameters of the dynamic model are identified by finding the optimal parameter minimizing the prediction error metric. This method is partly based on the previous study \cite{Wakita2022}. The coordinate systems and state variables were defined in \Cref{subsec:coo}. Then, \Cref{subsec:opt} describe the identification method. 
        
        \subsection{Coordinate systems}\label{subsec:coo}
\begin{figure}[t]
                \centering
                \includegraphics[width=0.8\linewidth]{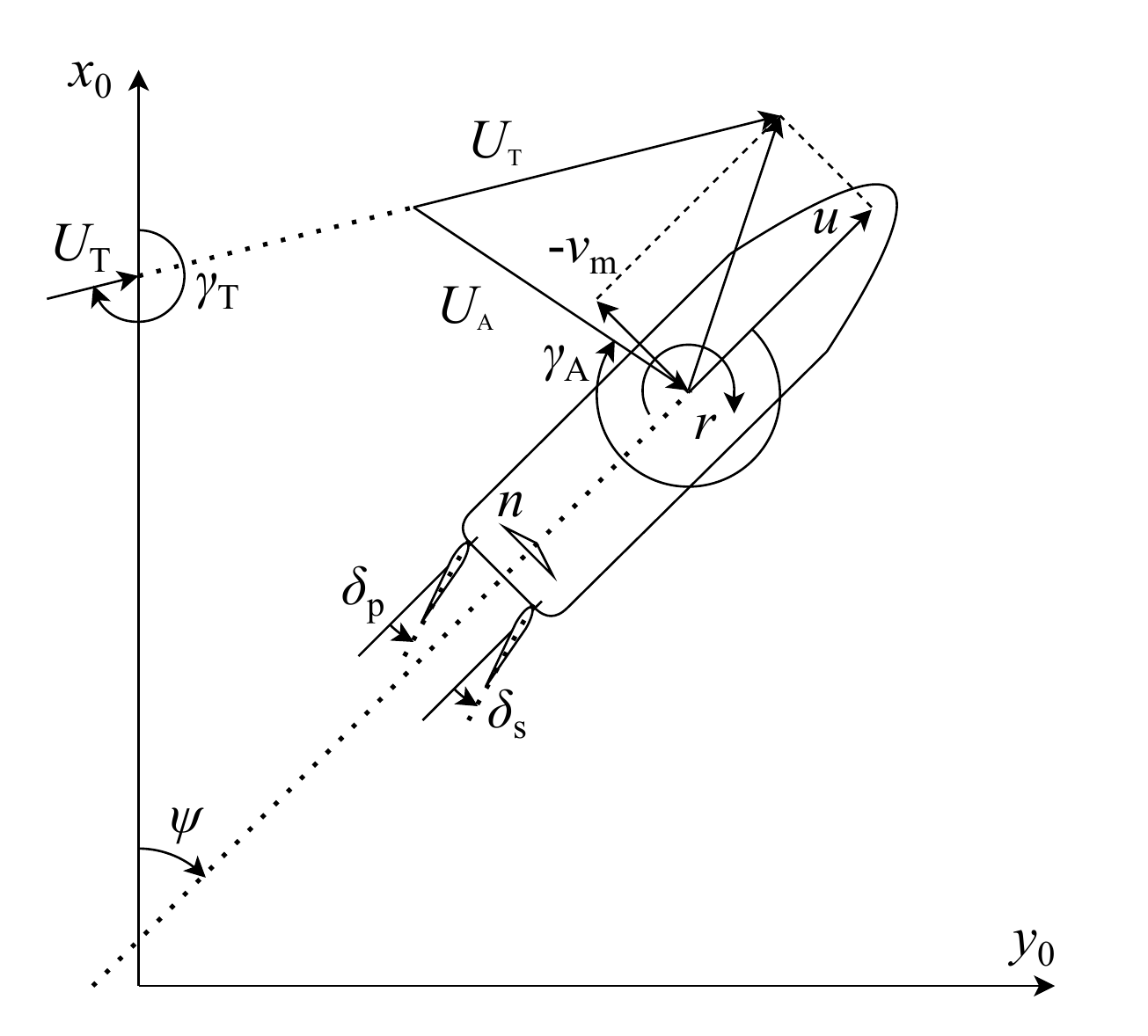}
                \caption{Coordinate systems.}
                \label{fig:coordinates}
            \end{figure}

\begin{figure}[t]
                \centering
                \includegraphics[width=0.9\linewidth]{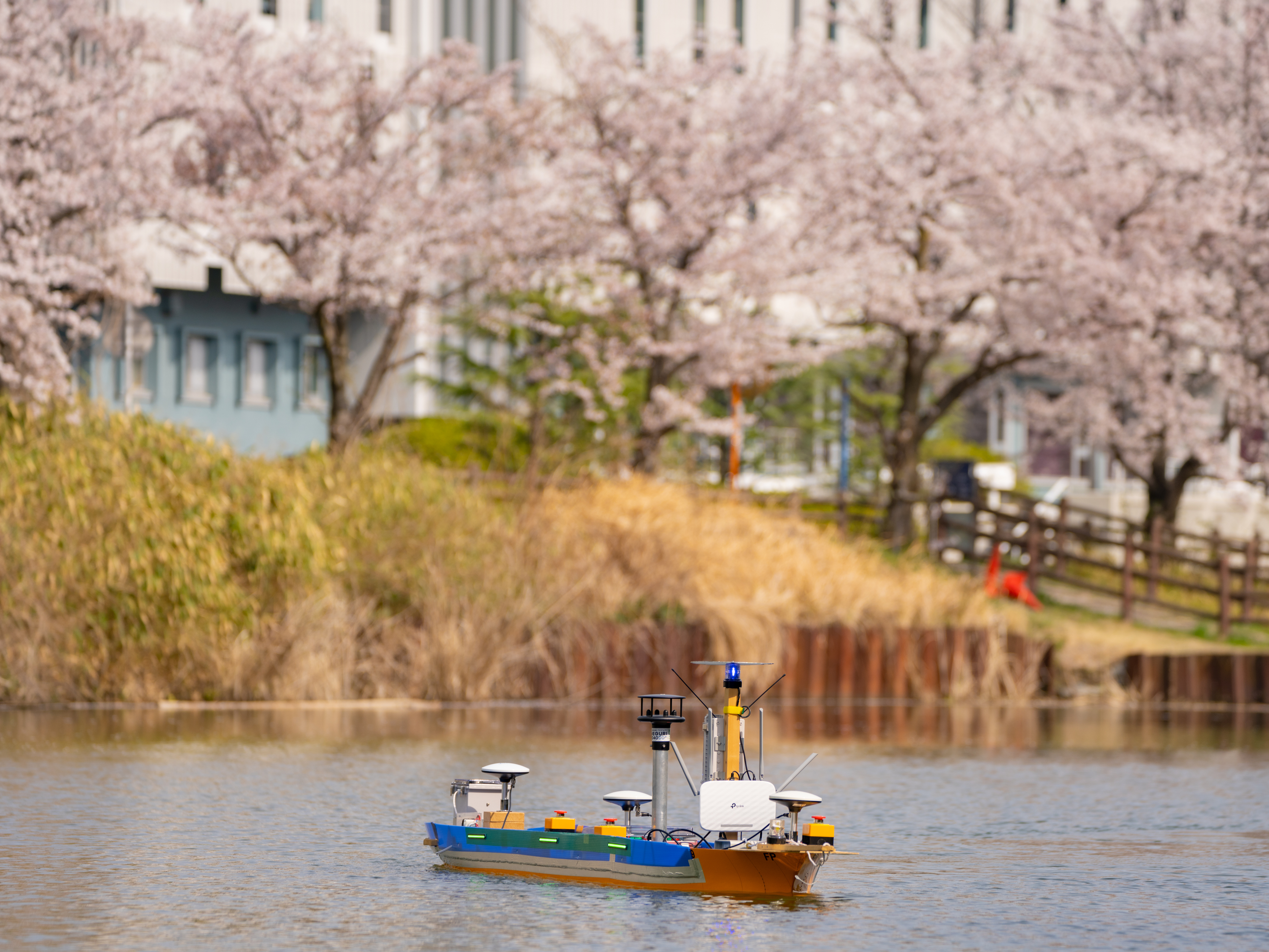}
                \caption{Subject model ship at the experimental pond.}
                \label{fig:model_ship}
            \end{figure}

            The earth-fixed coordinate system $\mathrm{O}-x_{0}y_{0}$ and a ship-fixed coordinate system $\mathrm{O}-xy$ are defined as \Cref{fig:coordinates}. Note that the origin of the coordinate system $\mathrm{O}-xy$ is fixed at the midship. Then, the subject ship of this study is the model ship equipped with a single-propeller, VecTwin rudder, and a bow thruster, and is shown in \Cref{fig:model_ship}. The port and starboard side rudder angle is defined as $\delta_{\mathrm{p}}$ and $\delta_{\mathrm{s}}$, respectively. The propeller revolution number is $n_{\mathrm{p}}$. These actuator states are defined as $\boldsymbol{u} \equiv \left(\delta_{\mathrm{p}}, \delta_{\mathrm{s}}, n_{\mathrm{p}}\right)^{\mathsf{T}} \in \mathbb{R}^{3}$. Note that this study used a bow thruster revolution number of zero.
            
            The heading angle from the $x_{0}$ axis is $\psi$. The surge, sway (at midship), and yaw angle velocity are denoted as $u$, $v_{\mathrm{m}}$, and $r$, respectively. These ship state variables are defined as $\boldsymbol{x} \equiv \left( x_{0}, u, y _{0}, v_{\mathrm{m}}, \psi, r \right)^{\mathsf{T}} \in \mathbb{R}^6$. For the convenience of explanation, the ship position and heading angle are defined as  $\boldsymbol{\eta} \equiv \left(x_{0}, y_{0}, \psi \right)^{\mathsf{T}} \in \mathbb{R}^{3}$ and the surge, sway, and yaw angle velocity are defined as $\boldsymbol{\nu} \equiv \left(u, v_{\mathrm{m}}, r\right)^{\mathsf{T}} \in \mathbb{R}^{3}$. The true wind speed and wind direction are defined as $U_{\mathrm{T}}$ and $\gamma_{\mathrm{T}}$, respectively. The apparent wind speed and direction are $U_{\mathrm{A}}$ and $\gamma_{\mathrm{A}}$, respectively. The apparent wind states are defined as  $\boldsymbol{\omega} = (U_{\mathrm{A}},\ \gamma_{\mathrm{A}})^\mathsf{T} \in \mathbb{R}^{2}$. This study assumes that the wind speed and direction are uniform in space and depend on the physical time.
            
        \subsection{Optimization problem}\label{subsec:opt}
            The parameter identification method used in this study was formulated as a minimization problem of the prediction error metric of the dynamic model. The prediction error was calculated by comparing the measured state variables and the state variable simulated by the dynamic model. The method was described in the remainder of this section.
            
            Let us assume that the time series of ship state variables $\boldsymbol{x}$ and actuator state variables $\boldsymbol{u}$, apparent wind state variables $\boldsymbol{\omega}$ were given, and $N$ denote the number of time series. This dataset is defined as follows:
            \begin{equation}
                \mathcal{D} \equiv \Bigl\{\bigl(\boldsymbol{x}_{n}\left(t\right), \boldsymbol{u}_{n}\left(t\right), \boldsymbol{\omega}_{n}\left(t\right) \mid t \in [t_{0, n}, t_{1, n}] \bigr)_{}\Bigr\}_{n=1, 2, \ldots, N} \enspace.
                \label{eq:countinuousdataset}
            \end{equation}
            Here, the subscript $n$ indicates that it is the $n$-th time series data.
            The effect of waves is not considered in this study and currents are ignored for simplicity since this study focuses on the harbor, berthing, and unberthing maneuver. The dynamic model is expressed as follows:
            \begin{equation}
                \dot{\boldsymbol{x}}(t)=\boldsymbol{f}\left(\boldsymbol{x}(t), \boldsymbol{u}(t), \boldsymbol{\omega}(t) ; \boldsymbol{\theta}\right) \enspace.
                \label{eq:ode}
            \end{equation}
            Here, $\boldsymbol{\theta}$ represents the parameters included in the dynamic model.
            
            The ship state variables can be simulated with the use of a dynamic model and the given dataset. The simulated ship state variables $\boldsymbol{x}^{\mathrm{sim}}_{n}$ is expressed as follows:
            \begin{equation}
                \begin{aligned}
                    \boldsymbol{x}^{\mathrm{sim}}_{n}(t; \boldsymbol{\theta}) = & \ \boldsymbol{x}_{n}(t_{0, n}) \\
                    & + \int^{t}_{t_{0, n}}\boldsymbol{f}\left(\boldsymbol{x}^{\mathrm{sim}}_{n}(\tau; \boldsymbol{\theta}), \boldsymbol{u}_{n}(\tau), \boldsymbol{\omega}_{n}(\tau) ; \boldsymbol{\theta}\right) \mathrm{d}\tau \\
                    & \quad \quad \quad \quad \quad \quad \quad \quad \text{for} \quad n = 1, 2, \ldots, N \enspace.
                \end{aligned}    
                \label{eq:continuousdynamics}
            \end{equation}
            The prediction error of the dynamic model is defined by comparing these simulated ship state variables $\boldsymbol{x}^{\mathrm{sim}}_{n}$ with that of the dataset $\boldsymbol{x}_{n}$. The evaluation function of the prediction error is defined as follows:
            \begin{equation}
                \mathcal{L}\left(\boldsymbol{\theta} ; \mathcal{D}\right) = \frac{1}{N}\sum_{n=1}^{N} \left\{ \int^{t_{1, n}}_{t_{0, n}} d\left(\boldsymbol{x}^{\mathrm{sim}}_{n}(t; \boldsymbol{\theta}), \boldsymbol{x}_{n}(t)\right) \mathrm{d}t \right\} \enspace,
                \label{eq:continuousobj}
            \end{equation}
            where, $d\left(\boldsymbol{x}^{\mathrm{sim}}, \boldsymbol{x}\right)$ is a function that returns the error between $\boldsymbol{x}^{\mathrm{sim}}$ and $\boldsymbol{x}$ as a scalar value, and defined as follows:
            \begin{equation}
                d\left(\boldsymbol{x}^{\mathrm{sim}}, \boldsymbol{x}\right) = \left\|\boldsymbol{w} \cdot \left(\boldsymbol{x}^{\mathrm{sim}} -  \boldsymbol{x}\right)\right\|^{2} \enspace.
                \label{eq:diffmetric}
            \end{equation}
            Here, $\boldsymbol{w} \in \mathbb{R}^{6}$ is a weight vector. This weight vector was introduced to compensate for the scale differences of state variables. 
            Although the model parameters can be identified by minimizing the prediction error defined in \Cref{eq:continuousobj}, a regularization term was added to the optimization target to avoid overfitting. Thus, the model parameters identified in the dataset are represented as follows:
            \begin{equation}
                \boldsymbol{\theta}_{\mathrm{opt}} = \argmin \left\{ \mathcal{L}\left(\boldsymbol{\theta} ; \mathcal{D}\right) + \lambda \left\|\boldsymbol{\theta}\right\|^{2} \right\} \enspace,
                \label{eq:minobj}
            \end{equation}
            where $\lambda$ is the regularization parameter.

            Although the time series data of \Cref{eq:countinuousdataset} are defined as continuous time values, the actual measured time series data are given as discrete time values. Indeed, the dataset obtained from the full-scale ship trials or free-running model tests is expressed as follows:
            \begin{equation}
                \mathcal{D} \equiv \Bigl\{\bigl(\boldsymbol{x}_{n}\left(t_{i}\right), \boldsymbol{u}_{n}\left(t_{i}\right), \boldsymbol{\omega}_{n}\left(t_{i}\right)\bigr)_{i=0, 1, \ldots, I_{n}-1}\Bigr\}_{n=1, 2, \ldots, N} \enspace.
                \label{eq:discretedataset}
            \end{equation}
            Here, the $t_{i}$ represents the physical time of the $i$-th time step, and $I_{n}$ denotes the number of time steps in the $n$-th time series data.
            It is hard to calculate \Cref{eq:continuousobj} and \Cref{eq:continuousdynamics} analytically. Therefore, this study calculated the time integration of \Cref{eq:continuousobj} by trapezoidal approximation as follows: 
            \begin{equation}
                \mathcal{L}\left(\boldsymbol{\theta} ; \mathcal{D}\right) = \frac{1}{N}\sum_{n=1}^{N} \left\{ \sum_{i=0}^{I_{n}-2} \frac{d_{n,i+1}+d_{n,i}}{2} \Delta t_{i} \right\} \enspace,
                \label{eq:discreteobj}
            \end{equation}
            where, $d_{n,i}=d\left(\boldsymbol{x}^{\mathrm{sim}}_{n}(t_{i}; \boldsymbol{\theta}), \boldsymbol{x}_{n}(t_{i})\right)$, $\Delta t_{i} = t_{i+1}-t_{i}$.
            The simulated ship state variables $\boldsymbol{x}^{\mathrm{sim}}_{n}$ were obtained by using the Euler method, as follows:
            \begin{equation}
                \begin{aligned}
                  \MoveEqLeft \boldsymbol{x}^{\mathrm{sim}}_{n} \left(t_{i}; \boldsymbol{\theta}\right) = \boldsymbol{x}_{n}\left(t_{0}\right) \\
                  & + \sum_{k=0}^{i-1} \Delta t_{k} \cdot \boldsymbol{f}\left(\boldsymbol{x}^{\mathrm{sim}}_{n}\left(t_{k}; \boldsymbol{\theta}\right), \boldsymbol{u}_{n}\left(t_{k}\right), \boldsymbol{\omega}_{n}\left(t_{k}\right) ; \boldsymbol{\theta}\right) & \\
                  & \quad \quad \quad \quad \quad \quad \quad \quad \quad \quad \quad \text{for} \quad n = 1, 2, \ldots, N \enspace.
                \end{aligned}
                \label{eq:discretedynamics}
            \end{equation}
            \Cref{eq:discreteobj,eq:discretedynamics} can be easily calculated numerically.
        
    \section{Data Augmentation Methods}\label{sec:aug}
        This section describes the data augmentation methods applied to the identification problem described in \Cref{sec:problem}. 
        
        Data augmentation generates synthetic data by transforming the original dataset. In this study, we focus on magnitude and time domain transformation-based data augmentation methods introduced in the previous study \cite{Iwana2021}. These methods are performed by transforming values or the time axis of time series data and include jittering, rotation, scaling, magnitude warping, slicing, permutation, and time warping. However, not all of these methods can be applied to the identification problem of \Cref{sec:problem}. For example, the window warping method distorts the meaning of the data because the time derivative of the state variables (the acceleration) is changed significantly.

        Among these methods, slicing \cite{Cui2016,leguennec2016} and jittering \cite{Bishop1995,Um2017,RASHID2019} are expected to be applicable. Slicing generates synthetic data by extracting slices from the original time series. Slicing does not alter the meaning of the original data, as it does not affect the time derivative. On the other hand, jittering generates synthetic data by adding noise to the time series data. Jittering can generate data that resembles a different realization of sensor data when noise, whose magnitude is close to the observation accuracy, is used. Jittering also does not significantly distort the meaning of the original data if the appropriate noise is used. Therefore, this study employs slicing and jittering.
        
        In this paper, the dataset without data augmentation is denoted as the reference dataset. The reference dataset is defined in \Cref{subsec:ref}, and the augmented dataset from \Cref{eq:discretedataset} by Jittering and/or slicing is expressed in \Cref{subsec:jit,subsec:sli,subsec:slijit}.
    
        \subsection{Definition of the reference dataset}\label{subsec:ref}
            The reference dataset without data augmentation is described here. The measured time series data do not necessarily have constant time steps. In other words, $I_n$ in \Cref{eq:discretedataset} is not necessarily constant for $n$. In this case, the numerical simulation requires a long computational time when the time step $I_n$ is large. Therefore, the time-series data are divided by a certain number of time steps, and the divided time-series dataset is defined as a reference dataset.
            
            The $n$-th time series data are divided at time step $I (< I_n)$. Then, the $k$-th time series data divided from the $n$-th time series data are expressed as follows:
            \begin{equation}
                \mathcal{D}^{(\text{ref})}_{n,k} \equiv \bigl(\boldsymbol{x}_{n}\left(t_{kI+i}\right), \boldsymbol{u}_{n}\left(t_{kI+i}\right), \boldsymbol{\omega}_{n}\left(t_{kI+i}\right)\bigr)_{i=0, \ldots, I-1} \enspace.
                \label{eq:basedata}
            \end{equation}
            Here, $K_{n}$ is the number of divisions and is denoted as follows:
            \begin{equation}
                K_{n} = \left\lfloor \frac{I_{n}}{I} \right\rfloor \enspace.
                \label{eq:baseKn}
            \end{equation}
            Note that $\lfloor \cdot \rfloor$ denotes the floor function.
            Thus, the reference dataset can be obtained by dividing $N$ time series data, and is represented as follows:
            \begin{equation}
                \mathcal{D}^{(\text{ref})} \equiv \left\{\left\{\mathcal{D}^{(\text{ref})}_{n,k}\right\}_{k=0, 1, \ldots, K_{n}-1}\right\}_{n=1, 2, \ldots, N} \enspace.
                \label{eq:basedataset}
            \end{equation}
        
        \subsection{Data augmentation by slicing}\label{subsec:sli}
            Slicing generates synthetic data by extracting the data with a certain time step from the original time-series data. The extracted data are selected randomly \cite{leguennec2016} or by sliding the start time step \cite{Cui2016}. This study used the latter method. 
            
            In this paper, the time step of the extracted time series data is equivalent to $I$, and the start time step is defined as $\varphi$. Then, the extracted time series data are expressed as follows:
            \begin{equation}
                \mathcal{D}^{(\text{sli})}_{n,\varphi} \equiv \Bigl(\boldsymbol{x}_{n}\left(t_{\varphi+i}\right), \boldsymbol{u}_{n}\left(t_{\varphi+i}\right), \boldsymbol{\omega}_{n}\left(t_{\varphi+i}\right)\Bigr)_{i=0, \ldots, I-1} \enspace.
                \label{eq:slidata}
            \end{equation}
            Here, $0 \le \varphi \le I_n - I$ must be satisfied. In this study, we extract data by sliding $\varphi$ at a step interval $S$. Thus, the dataset augmented by slicing is defined as follows:
            \begin{equation}
                \mathcal{D}^{(\text{sli})} \equiv \left\{\left\{\mathcal{D}^{(\text{sli})}_{n,\varphi}\right\}_{\varphi=0, S, \ldots, \left\lfloor \frac{I_n - I}{S} \right\rfloor S}\right\}_{n=1, 2, \ldots, N} \enspace.
                \label{eq:slidataset}
            \end{equation}

            In the parameter identification method described in \Cref{sec:problem}, the estimation accuracy of the dynamic model is dependent on the accuracy of the initial ship state variables $\boldsymbol{x}_{n}\left(t_{0}\right)$. The observation error of $\boldsymbol{x}_{n}\left(t_{0}\right)$ affects the estimation accuracy. Therefore, applying slicing to the parameter identification is expected to reduce the effect of the specific observation error.

        \subsection{Data augmentation by jittering}\label{subsec:jit}
            Jittering generates synthetic data by adding noise to the time series data. In NNs, it is well known that adding noise to model inputs prevents overfitting and improves generalization performance \cite{Bishop1995}. In addition, jittering could be applied to sensor data \cite{Um2017,RASHID2019}, since this method assumes that time-series data contain noise.

            Normal noise independent in time and space is added to the ship state variables, and not added to the actuator and wind state variables. The $m$-th normal noise vector added to the ship state variables $\boldsymbol{x}_{n}\left(t_{i}\right)$ is defined as $\boldsymbol{\epsilon}_{n,m}\left(t_{i}\right) \sim \mathcal{N}\left(\boldsymbol{0}, \boldsymbol{\Sigma}_{\boldsymbol{x}}\right)$, where $\boldsymbol{\Sigma}_{\boldsymbol{x}}$ denotes the covariance matrix of the normal noise. The time series data to which noise is added is represented as follows:
            \begin{equation}
                \begin{aligned}
                    \mathcal{D}^{(\text{jit})}_{n,k,m} \equiv \bigl(\boldsymbol{x}_{n}\left(t_{kI+i}\right) + \boldsymbol{\epsilon}_{n,m}\left(t_{kI+i}\right), \quad\quad\quad\quad \quad\quad\quad\quad \\
                    \boldsymbol{u}_{n}\left(t_{kI+i}\right), \boldsymbol{\omega}_{n}\left(t_{kI+i}\right)\bigl)_{i=0, \ldots, I-1} \enspace. 
                \end{aligned} 
                \label{eq:jitdata}
            \end{equation}
            Thus, the dataset augmented by jittering is defined as follows:
            \begin{equation}
                \mathcal{D}^{(\text{jit})} \equiv \left\{\left\{ \mathcal{D}^{(\text{jit})}_{n,k,m} \right\}_{k=0, 1, \ldots, K_{n}-1} \right\}_{n=1, 2, \ldots, N, m=1, 2, \ldots, M} \enspace. 
                \label{eq:jitdataset}
            \end{equation}
            Jittering can increase the amount of data in proportion to the number of noises it generates. However, unnecessarily large noise deteriorates estimation accuracy. Therefore, the covariance matrix $\boldsymbol{\Sigma}_{\boldsymbol{x}}$ of added noise was determined according to the observation accuracy of the measurement equipment.
            
        \subsection{Data augmentation by slicing and jittering}\label{subsec:slijit}
            The data augmentation method that uses slicing and jittering simultaneously is not so difficult to accomplish. The synthetic data can be generated by adding noise to the time series data of \Cref{eq:slidata}. The generated time series data are expressed as follows:
            \begin{equation}
                \begin{aligned}
                    \mathcal{D}^{(\text{sli}\times\text{jit})}_{n,\varphi,m} \equiv \Bigl(\boldsymbol{x}_{n}\left(t_{\varphi+i}\right) + \boldsymbol{\epsilon}_{n,m}\left(t_{\varphi+i}\right), \boldsymbol{u}_{n}\left(t_{\varphi+i}\right),  \\
                    \boldsymbol{\omega}_{n}\left(t_{\varphi+i}\right)\Bigr)_{i=0, \ldots, I-1} \enspace. 
                \end{aligned}
                \label{eq:slijitdata}
            \end{equation}
            Thus, the dataset augmented by slicing and Jittering is defined as follows:
            \begin{equation}
                \mathcal{D}^{(\text{sli}\times\text{jit})} \equiv \left\{ \left\{\mathcal{D}^{(\text{sli}\times\text{jit})}_{n,\varphi,m}\right\}_{\varphi=0, S, \ldots, \left\lfloor \frac{I_n - I}{S} \right\rfloor S} \right\}_{n=1, 2, \ldots, N, m=1, 2, \ldots, M} \enspace.
                \label{eq:slijitdataset}
            \end{equation}
        
    \section{Numerical Experiments}\label{sec:exp}

        One of the purposes of this paper is to demonstrate the effectiveness of the data augmentation methods described in \Cref{sec:aug}. Numerical experiments were conducted using datasets obtained from free-running model tests. In this section, the numerical experiments are described and the results are presented. \Cref{subsec:detaset} describes the prepared dataset, \Cref{subsec:model} describes the model used, \Cref{subsec:optimization} describes the optimization method and result, and \Cref{subsec:result} shows the prediction results of the identified dynamic model.

        \subsection{Dataset}\label{subsec:detaset}
            The free-running model tests were conducted and the time series data of the ship state variables $\boldsymbol{x}$, actuator state variables $\boldsymbol{u}$, and relative wind state variables $\boldsymbol{\omega}$ were measured. The configuration of the observation equipment of the model ship was the same as in the previous study \cite{Miyauchi2023}. The $x_{0}$ and $y_{0}$ were calculated by transforming the measured position by Global Navigation Satellite System (GNSS) to the midship position. The heading angle $\psi$ and yaw angle velocity $r$ were measured by Fiber Optical Gyro (FOG). $u$ and $v_{\mathrm{m}}$ were calculated from the speed over ground, the course over ground, and the heading angle measured by GNSS and FOG, respectively. The apparent wind speed $U_{\mathrm{A}}$ and direction $\gamma_{\mathrm{A}}$ were measured by an ultrasonic anemometer. Although the measurement frequency was 10 Hz, the data used for the parameter identification were downsampled to 1 Hz.
    
            The maneuvers conducted in free-running model tests were random maneuvers. In random maneuvers, control inputs are selected randomly by the human operator to collect datasets including various ship and actuator states. These model tests were conducted at Inukai Pond, which is an experimental pond at Osaka University. The details of these tests were described in the previous study \cite{Wakita2022,Miyauchi2022}. The upper and lower limits of the actuator state variables are presented in \Cref{tab:actlim}. The obtained trajectory data and their measurement times are shown in \Cref{tab:trajdata}.
    
            \begin{table}[h]
                \centering
                \caption{Limit of control inputs.}
                \begin{tabular}{cc}
                    \hline Variable & Maximum and minimum \\
                    \hline$n_{\mathrm{P}} \ (\mathrm{rps})$ & {$[0,12.5]$} \\
                    $\delta_{\mathrm{s}} \ (\mathrm{deg.})$ & {$[-35,105]$} \\
                    $\delta_{\mathrm{p}} \ (\mathrm{deg.})$ & {$[-105,35]$} \\
                    \hline
                \end{tabular}
                \label{tab:actlim}
            \end{table}
    
            \begin{table*}[h]
              \begin{minipage}[c]{0.3\hsize}
                \centering
                \caption{Trajectory data collected by free-running model tests. The sampling frequency is 1 Hz.}
                \begin{tabular}{ll}\hline
                  Trajectory & Duration \\ \hline \hline
                  No. 1 & 500.5 (s) \\
                  No. 2 & 1801.8 (s) \\
                  No. 3 & 500.5 (s) \\
                  No. 4 & 1801.8 (s) \\
                  No. 5 & 1201.2 (s) \\
                  No. 6 & 1201.2 (s) \\ \hline
                \end{tabular}
                \label{tab:trajdata}
              \end{minipage}
              \begin{minipage}[c]{0.7\hsize}
                \centering
                \caption{Prepared dataset. Eight different training datasets are prepared.}
                 \begin{tabular}{l|ll}\hline
                    Dataset Name & Trajectory No. & Augmentation methods  \\ \hline\hline
                    $\mathcal{D}^{(\text{ref})}$ & No. 1 and 2 & \Cref{subsec:ref} ($I=100$) \\
                    $\mathcal{D}^{(\text{sli2})}$ & No. 1 and 2 & \Cref{subsec:sli} ($I=100, S=50$) \\
                    $\mathcal{D}^{(\text{sli10})}$ & No. 1 and 2 & \Cref{subsec:sli} ($I=100, S=10$) \\
                    $\mathcal{D}^{(\text{jit2})}$ & No. 1 and 2 & \Cref{subsec:jit} ($I=100, M=2$) \\
                    $\mathcal{D}^{(\text{jit10})}$ & No. 1 and 2 & \Cref{subsec:jit} ($I=100, M=10$) \\
                    $\mathcal{D}^{(\text{sli2}\times\text{jit2})}$ & No. 1 and 2 & \Cref{subsec:slijit} ($I=100, S=50, M=2$) \\
                    $\mathcal{D}^{(\text{sli10}\times\text{jit10})}$ & No. 1 and 2 & \Cref{subsec:slijit} ($I=100, S=10, M=10$) \\ 
                    $\mathcal{D}^{(\text{d-ref})}$ & No. 1, 2, 3 and 4 & \Cref{subsec:ref} ($I=100$) \\
                    \hline
                    $\mathcal{D}^{(\text{validation})}$ & No. 5 & \Cref{subsec:ref} ($I=100$) \\
                    $\mathcal{D}^{(\text{test})}$ & No. 6 & \Cref{subsec:ref} ($I=100$) \\
                    \hline
                \end{tabular}
                \label{tab:dataset}
              \end{minipage}
            \end{table*}
        
            The trajectories from No. 1 to No. 4 were used as training data, the trajectory of No. 5 as validation data, and the trajectory of No. 6 as test data. We prepared eight different training datasets to show the effectiveness of the data augmentation methods. The prepared datasets and their augmentation methods are listed in \Cref{tab:dataset}. $\mathcal{D}^{(\text{ref})}$ is a reference dataset without data augmentation. $\mathcal{D}^{(\text{sli2})}$ and $\mathcal{D}^{(\text{jit2})}$ are datasets that have been augmented to twice the amount of data by each data augmentation method, and $\mathcal{D}^{(\text{sli10})}$ and $\mathcal{D}^{(\text{jit10})}$ are datasets that have been augmented to 10 times the amount of data. $\mathcal{D}^{(\text{sli2}\times\text{jit2})}$ and $\mathcal{D}^{(\text{sli10}\times\text{jit10})}$ are augmented to 4 and 100 times the amount of data by both slicing and jittering, respectively. $\mathcal{D}^{(\text{d-ref})}$ has twice as much data without data augmentation. 

\begin{figure}[t]
                \centering
                \includegraphics[width=\linewidth]{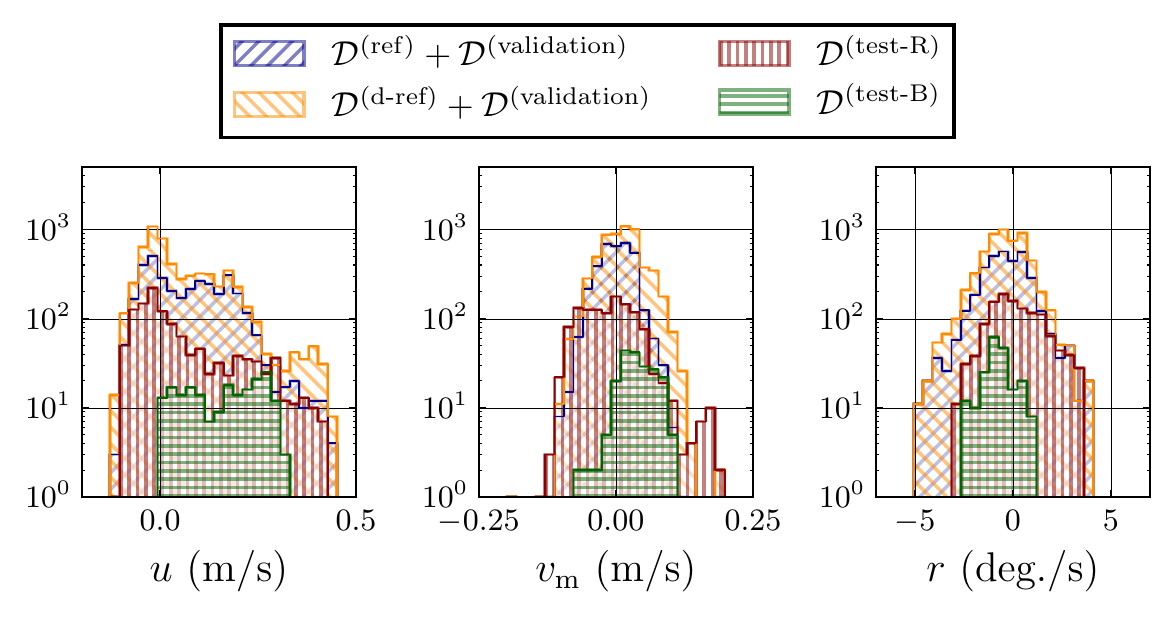}
                \caption{Histograms of ship and actuator state variables. Note that the vertical axes, which show the frequency, are scaled logarithmically.}
                \label{fig:single_histgram}
            \end{figure}

\begin{figure}[t]
                \centering
                \includegraphics[width=\linewidth]{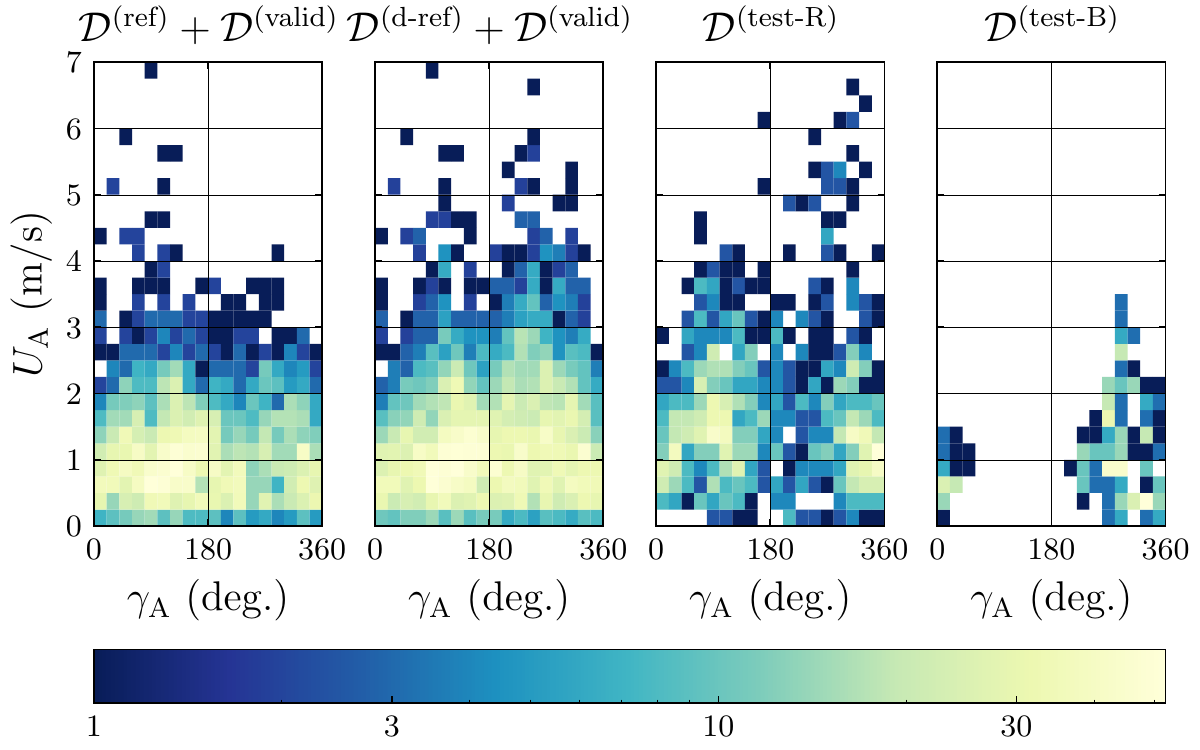}
                \caption{2D histogram of apparent wind speed and direction. The bin width of apparent wind speed is $0.25$ m/s, and that of apparent wind direction is $20$ degrees. The color bar shows the frequency with a logarithmic scale.}
                \label{fig:wind_histgram}
            \end{figure}

            To show the details of the prepared datasets, the histogram of the ship and actuator state variables $\boldsymbol{\nu}, \boldsymbol{u}$ for $\mathcal{D}^{(\text{ref})}$, $\mathcal{D}^{(\text{d-ref})}$, $\mathcal{D}^{(\text{validation})}$, and $\mathcal{D}^{(\text{test})}$ are presented in \Cref{fig:single_histgram}, and the 2D histogram of apparent wind speed and direction are presented in \Cref{fig:wind_histgram}. 
            These histograms shows that neither $\mathcal{D}^{(\text{ref})}$ nor $\mathcal{D}^{(\text{d-ref})}$ cover all data included in $\mathcal{D}^{(\text{test})}$. For instance, $\mathcal{D}^{(\text{ref})}$ does not include surge velocities of above $3.8$ m/s existing in $\mathcal{D}^{(\text{test})}$. Although $\mathcal{D}^{(\text{d-ref})}$ has more amount of data and a wider distribution than $\mathcal{D}^{(\text{ref})}$, $\mathcal{D}^{(\text{d-ref})}$ also does not include sway velocities of above $1.5$ m/s existing in $\mathcal{D}^{(\text{test})}$. Moreover, $\mathcal{D}^{(\text{test})}$ includes the wind state variables of around $U_{\mathrm{A}}=5.0 \ \mathrm{(m/s)}, \gamma_{\mathrm{A}}=300 \ \mathrm{(deg.)}$, while $\mathcal{D}^{(\text{ref})}$ does not.
            
            In free-running model tests, it is possible to obtain the dataset that completely covers the test dataset. However, in full-scale trials, it is difficult to do so because of the high costs and its time-consuming nature. In particular, wind disturbances cannot be controlled and large wind speeds rarely occur. In practical use, all possible state variables cannot be included in the training data, and there is a large possibility of encountering state variables that are not included in the training data. Therefore, this study shows the generalization performance for extrapolated state variables not included in the training data.

        \subsection{Mathematical Model}\label{subsec:model}
            In this study, the dynamic model is expressed using an NN \cite{rumelhart1985}. The NN model is a mathematical model that is inspired by the human brain. This model is known as the model that has high approximation capabilities \cite{Cybenko1989,HORNIK1991251}. In this study, an NN model consisting of only a fully connected layer was used. This model is often called the feedforward NN.
            
            The dynamic model defined in \Cref{eq:ode} is not directly expressed by NN. The time derivative of the surge, sway, and yaw angle velocity vector $\dot{\boldsymbol{\nu}}$ are predicted by the NN model, while that of the ship position and heading angle $\dot{\boldsymbol{\eta}}$ are expressed as follows:
            \begin{equation}
                \dot{\boldsymbol{\eta}}=
                    \begin{bmatrix}
                        \cos \psi & - \sin \psi & 0 \\
                        \sin \psi & \cos \psi & 0 \\
                        0 & 0 & 1 \\
                    \end{bmatrix} \boldsymbol{v} \enspace.
                \label{eq:kinema}
            \end{equation}

            Scale differences in NN inputs are likely to make the optimization problem challenging. Therefore, the NN inputs were standardized using the mean and standard deviation of the training data. The standardized variables of $\boldsymbol{\nu}, \boldsymbol{u}, \boldsymbol{\omega}$ are defined as $\bar{\boldsymbol{\nu}}, \bar{\boldsymbol{u}}, \bar{\boldsymbol{\omega}}$, respectively. The $j$-th components of these standardized variables are represented as follows:
            \begin{equation}
                \begin{aligned}
                    \bar{\nu}_{j} &= \left(\nu_{j}-\mu^{(\text{train})}_{\nu,j}\right)/\left(\sigma^{(\text{train})}_{\nu,j}\right) \\
                    \bar{u}_{j} &= \left(u_{j}-\mu^{(\text{train})}_{u,j}\right)/\left(\sigma^{(\text{train})}_{u,j}\right) \\
                    \bar{w}_{j} &= \left(w_{j}-\mu^{(\text{train})}_{w,j}\right)/\left(\sigma^{(\text{train})}_{w,j}\right) \enspace. \\ 
                \end{aligned}
                \label{eq:stdinput}
            \end{equation}
            In this study, the mean and standard deviation are calculated from $\mathcal{D}^{(\text{ref})}$ for all cases. Then, the NN inputs are defined as $\boldsymbol{s} = \left(\bar{\boldsymbol{\nu}}^{\mathsf{T}}, \bar{\boldsymbol{u}}^{\mathsf{T}}, \bar{\boldsymbol{\omega}}^{\mathsf{T}}\right)^{\mathsf{T}} \in \mathsf{R}^8$
    
            Furthermore, the NN outputs are also assumed to be standardized variables. The NN outputs are defined as $\boldsymbol{y} \in \mathbb{R}^3$, and then the $j$-th components of the predicted vector $\dot{\boldsymbol{\nu}}$ can be represented as follows:
            \begin{equation}
                \dot{v}_{j} = \sigma^{(\text{train})}_{\text{acc},j} \cdot y_{j} + \mu^{(\text{train})}_{\text{acc},j} \enspace,
                \label{eq:stdoutput}
            \end{equation}
            where $\mu^{(\text{train})}_{\text{acc},j}$ and $\sigma^{(\text{train})}_{\text{acc},j}$ denotes the $j$-th components of the mean and standard deviation, respectively, calculated from $\mathcal{D}^{(\text{ref})}$. Note that the variables of $\dot{\boldsymbol{\nu}}$ were calculated by numerical differentiation of $\boldsymbol{\nu}$ since those variables were not measured. 

            Therefore, it is the following function $\boldsymbol{f}_{\mathrm{NN}}$, not $\boldsymbol{f}$ in \Cref{eq:nnfunc}, that is approximated by the NN model:
            \begin{equation}
                \boldsymbol{y} = \boldsymbol{f}_{\mathrm{NN}}\left(\boldsymbol{s}; \boldsymbol{\theta}\right)
                \label{eq:nnfunc}
            \end{equation}
            The details of the NN model representing $\boldsymbol{f}_{\mathrm{NN}}$ are shown in \Cref{tab:networkparam}.
            
            \begin{table}[t]
                \centering
                \caption{Details of the NN model.}
                \begin{tabular}{c|ll}\hline
                    & Units of layer & Activation function \\ \hline\hline
                    Input    & 8     & tanh \\
                    Middle 1 & 256   & tanh \\
                    Middle 2 & 256   & tanh \\
                    Middle 3 & 256   & tanh \\
                    Middle 4 & 256   & tanh \\
                    Output   & 3     & linear \\ \hline
                \end{tabular}
                \label{tab:networkparam}
            \end{table}

        \subsection{Optimization Methods and Results}\label{subsec:optimization}

            \begin{table}[t]
                \centering
                \caption{Hyperparameters in training.}
                \begin{tabular}{ll}\hline
                    Learning rate & $1.0\times10^{-4}$ \\
                    $\lambda$ (\Cref{eq:minobj}) & $1.0\times10^{-2}$ \\
                    $\boldsymbol{w}$ (\Cref{eq:diffmetric}) & $(0, 100, 0, 100, 0, 10)^{\mathsf{T}}$ \\
                    $\Sigma_{\boldsymbol{x}}$ & $\mathrm{diag}(0.0, 0.01^2, 0.0, 0.01^2, 0.0, 0.1^2)$ \\
                    \hline
                \end{tabular}
                \label{tab:hyperparam}
            \end{table}
            
            \begin{figure}[t]
                \centering
                \includegraphics[width=\linewidth]{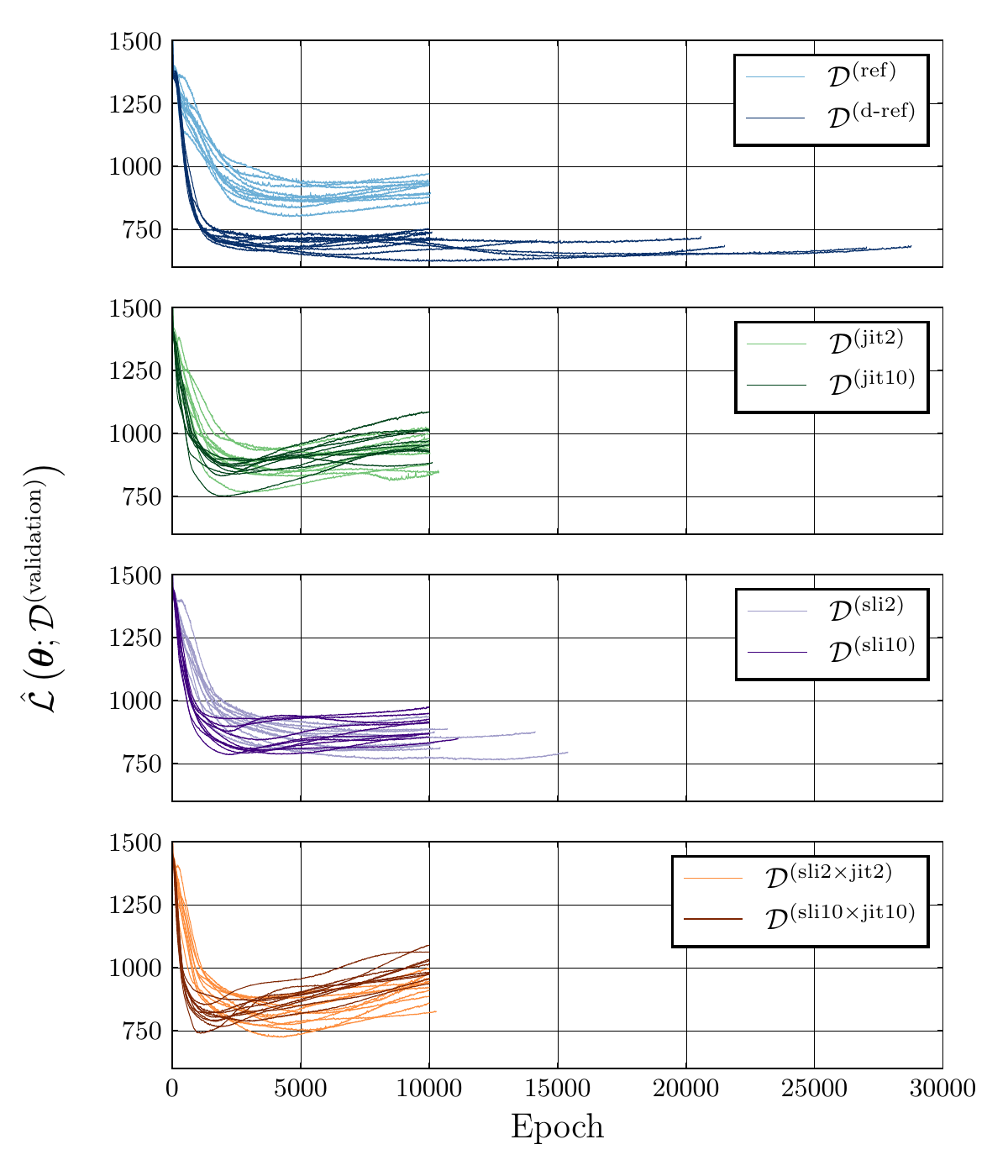}
                \caption{Exponential moving average values of the evaluation function in the validation dataset. The legend implies the used training dataset. Ten training results with different random numbers for each dataset are presented.}
                \label{fig:valid_loss}
            \end{figure}
            
            The optimization method of the NN model is described, and the optimization results are presented.
            The parameter of the NN model is optimized by a gradient descent-based optimization method, Adam \cite{kingma2014}, and Pytorch, an open-source Python library for machine learning, was used for the implementation. In optimizing the NN model, the training dataset was randomly split into three subsets, and we compute the gradients and update parameters for each subset. The number of cycles that pass through a training dataset is called an epoch.
            
            The hyperparameters used in training are presented in \Cref{tab:hyperparam}. Note that the 1, 3, and 5 components of $\boldsymbol{w}$ are set to zero, to ignore the errors of the position and heading angle. The covariance matrix $\Sigma_{\boldsymbol{x}}$ is determined so that noise is added only to the velocity and angular velocity. 
            
            The optimization was conducted 10 times for each dataset presented in \Cref{tab:dataset} by changing the random number. The optimal parameter was computed for each training dataset and each random number. The exponential moving average values of the evaluation function using the validation dataset at each epoch are presented in \Cref{fig:valid_loss}. Note that the exponential moving average values were calculated as follows:
            \begin{equation}
                \hat{\mathcal{L}}_{i} = \left\{\begin{array}{ll}
                        \alpha \mathcal{L}_{i} + \left(1-\alpha\right)\hat{\mathcal{L}}_{i-1} & (i \neq 0) \\
                        \mathcal{L}_{0} & (i = 0)
                    \end{array} \right.
                 \enspace,
                \label{eq:ema}
            \end{equation}
            where $\alpha = 0.1$, $\mathcal{L}_{i}$ means the evaluation function values of the $i$-th epoch, and $\hat{\mathcal{L}}_{i}$ means the exponential moving average ones.
            The training was terminated when the number of epochs exceeded $10,000$ and the value of the evaluation function values for the validation dataset satisfied the following inequality:
            \begin{equation}
                \mathcal{L}\left(\boldsymbol{\theta}\right) >
                0.1 \times \left( \mathcal{L}\left(\boldsymbol{\theta}_{\mathrm{init}} \right) - \mathcal{L}\left(\boldsymbol{\theta}_{\mathrm{min}} \right) \right) + \mathcal{L}\left(\boldsymbol{\theta}_{\mathrm{min}} \right) \enspace.
                \label{eq:stopcondi}
            \end{equation}
            Here, $\boldsymbol{\theta}_{\mathrm{init}}$ denotes the initial parameter and $\boldsymbol{\theta}_{\mathrm{min}}$ denotes the parameter with the smallest values during training. In \Cref{eq:stopcondi}, the used dataset $\mathcal{D}^{(\text{validation})}$ is omitted for simplicity. 
            
            \Cref{fig:valid_loss} shows that the training was terminated by satisfying \Cref{eq:stopcondi} in all cases. These results indicate that overfitting to the training dataset occurred.
            Therefore, in this study, the optimal parameters are the parameter with the smallest values of the evaluation function for the validation dataset to avoid overfitting and are denoted as $\boldsymbol{\theta}_{\mathrm{opt}}$.
            
        \subsection{Prediction Results}\label{subsec:result}

            \begin{figure}[t]
                \centering
                \includegraphics[width=\linewidth]{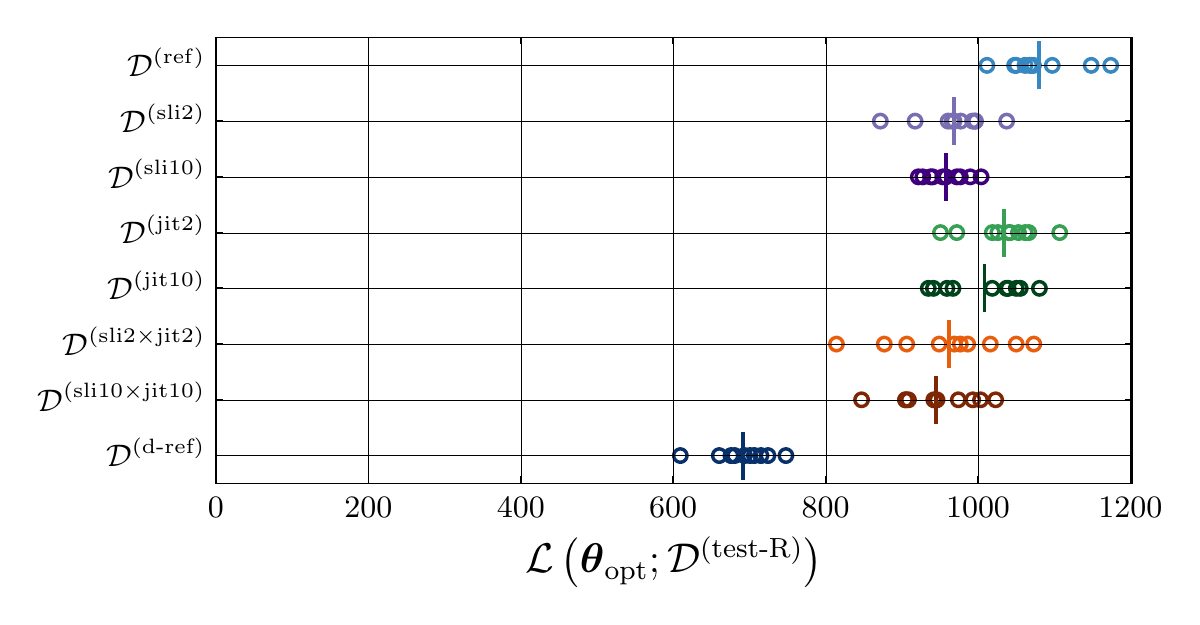}
                \caption{Evaluation function values in the test dataset. Vertical bars represent mean values.}
                \label{fig:test_loss}
            \end{figure}
            
            \begin{figure}[t]
                \centering
                \includegraphics[width=\linewidth]{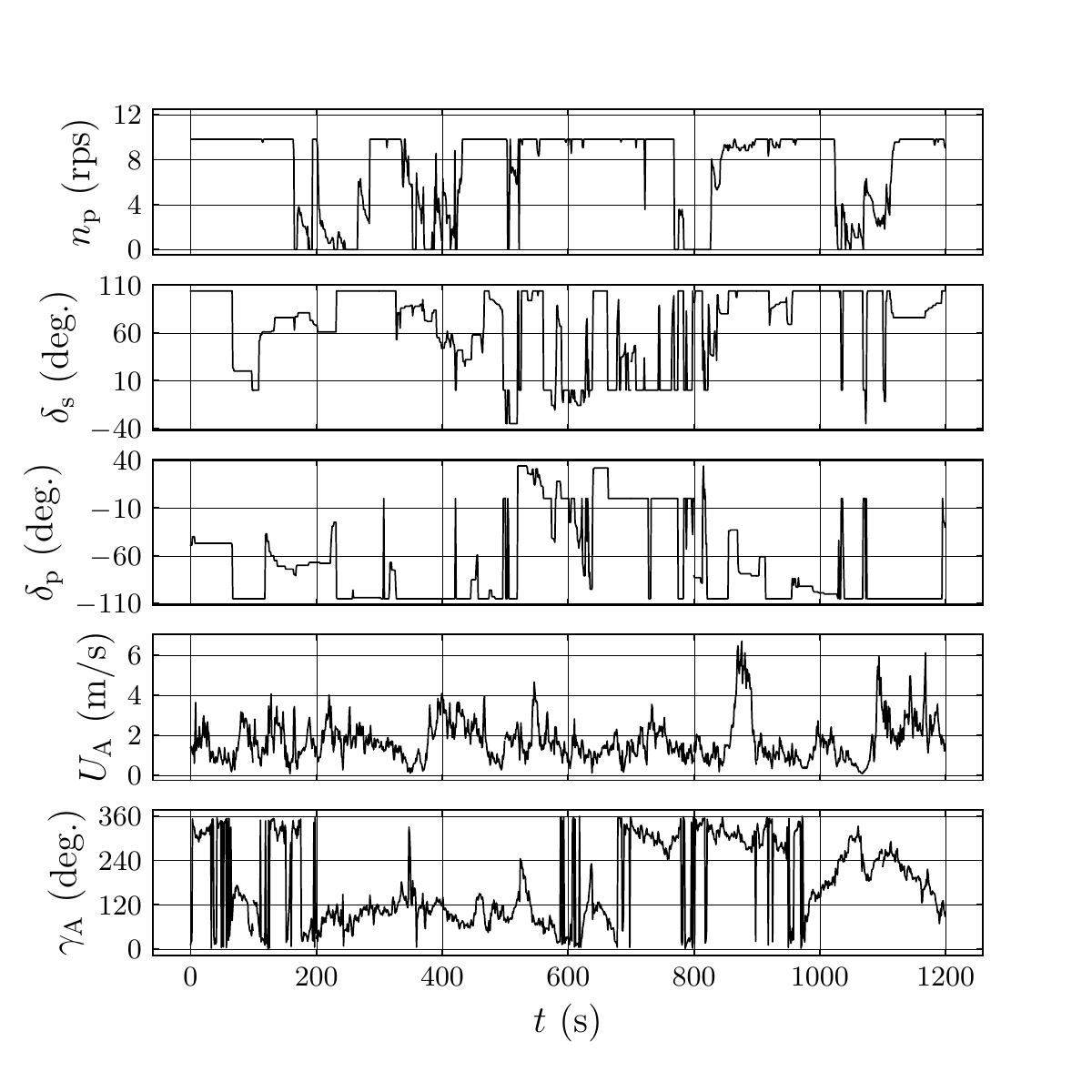}
                \caption{Time series data of control inputs $\boldsymbol{u}$ and apparent wind velocity and direction $\boldsymbol{\omega}$ in test data $\mathcal{D}^{(\text{test})}$.}
                \label{fig:traj_supplement}
            \end{figure}
    
            \begin{figure}[t]
                \centering
                \includegraphics[width=\linewidth]{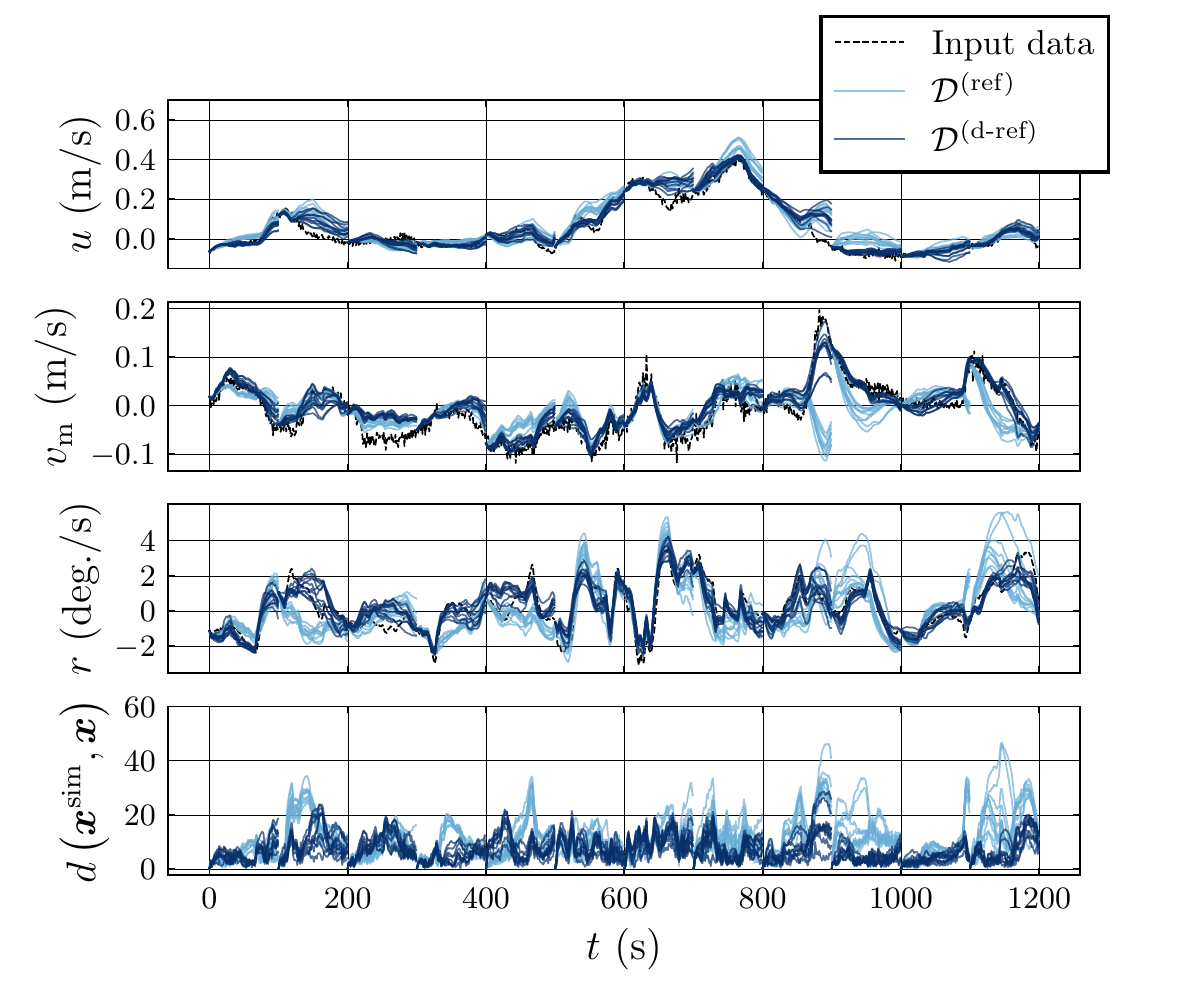}
                \caption{Prediction ship state variables $\boldsymbol{\nu}$ and error $d\left(\boldsymbol{x}^{\mathrm{sim}}, \boldsymbol{x}\right)$ using the optimal parameter trained by $\mathcal{D}^{(\text{ref})}$ and $\mathcal{D}^{(\text{d-ref})}$.}
                \label{fig:traj_base_double_error}
            \end{figure}
            \begin{figure}[t]
                \centering
                \includegraphics[width=\linewidth]{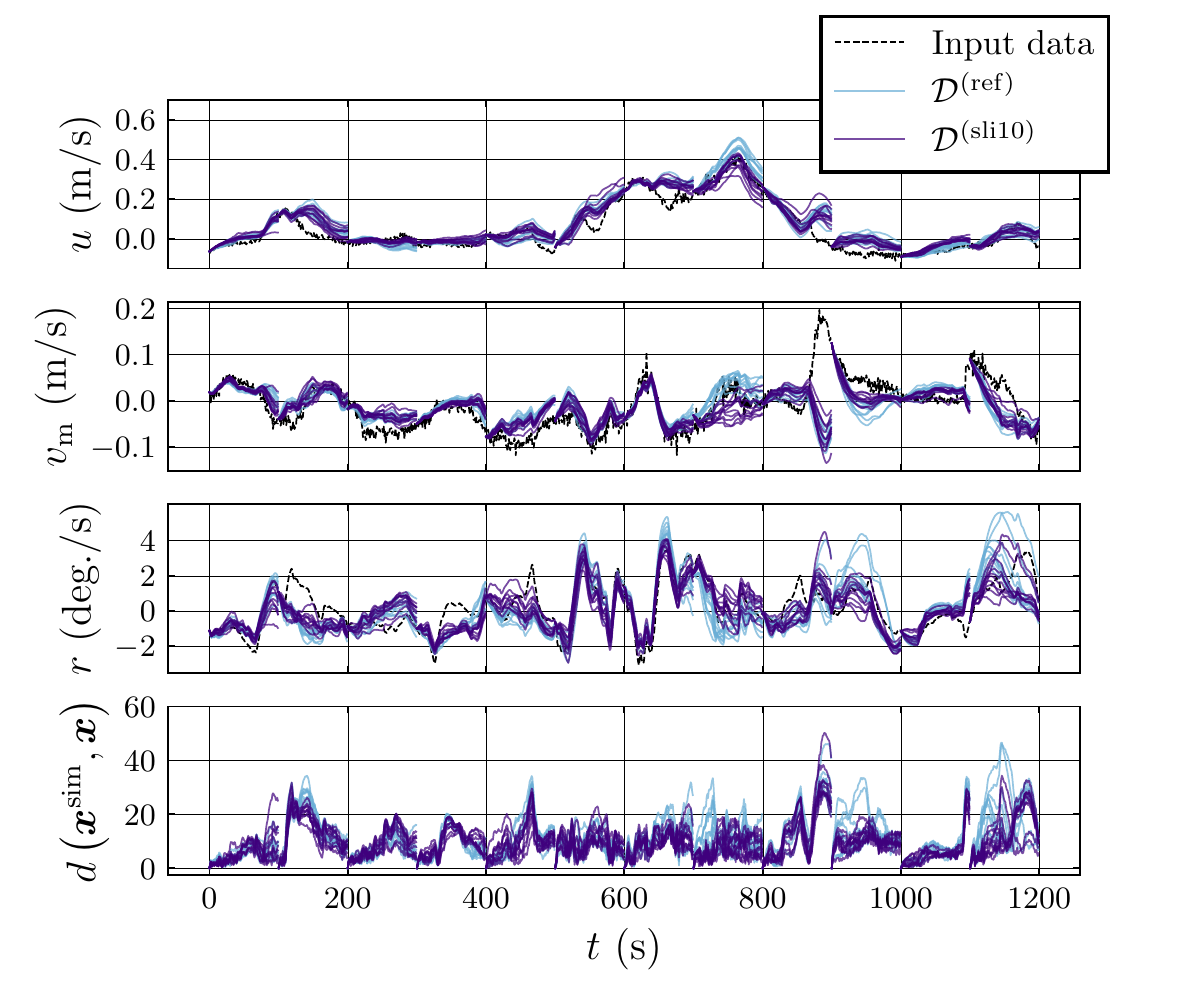}
                \caption{Prediction ship state variables $\boldsymbol{\nu}$ and error $d\left(\boldsymbol{x}^{\mathrm{sim}}, \boldsymbol{x}\right)$ using the optimal parameter trained by $\mathcal{D}^{(\text{ref})}$ and $\mathcal{D}^{(\text{sli10})}$.}
                \label{fig:traj_base_slicing10_error}
            \end{figure}
            \begin{figure}[t]
                \centering
                \includegraphics[width=\linewidth]{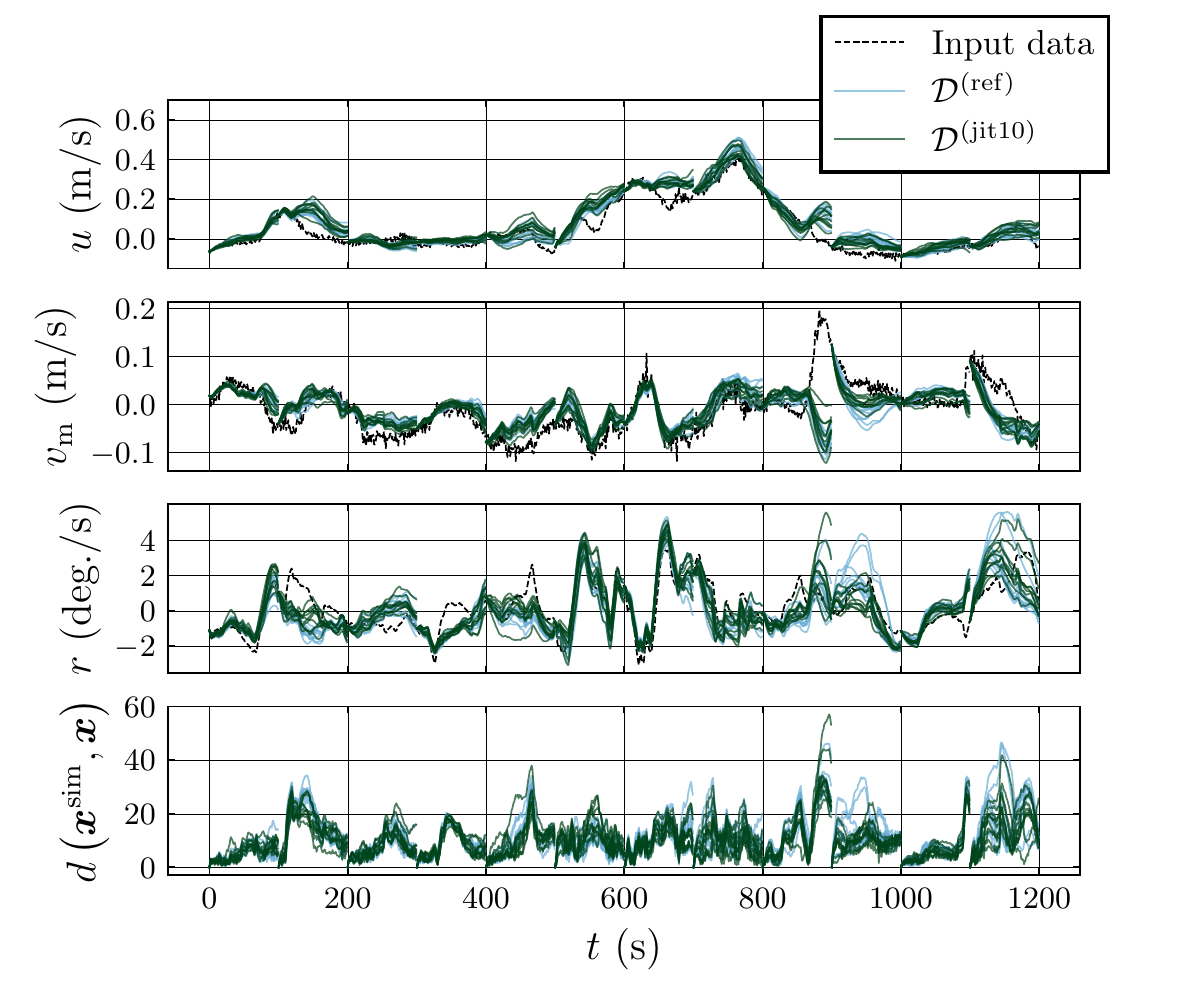}
                \caption{Prediction ship state variables $\boldsymbol{\nu}$ and error $d\left(\boldsymbol{x}^{\mathrm{sim}}, \boldsymbol{x}\right)$ using the optimal parameter trained by $\mathcal{D}^{(\text{ref})}$ and $\mathcal{D}^{(\text{jit10})}$.}
                \label{fig:traj_base_jittering10_error}
            \end{figure}
            \begin{figure}[t]
                \centering
                \includegraphics[width=\linewidth]{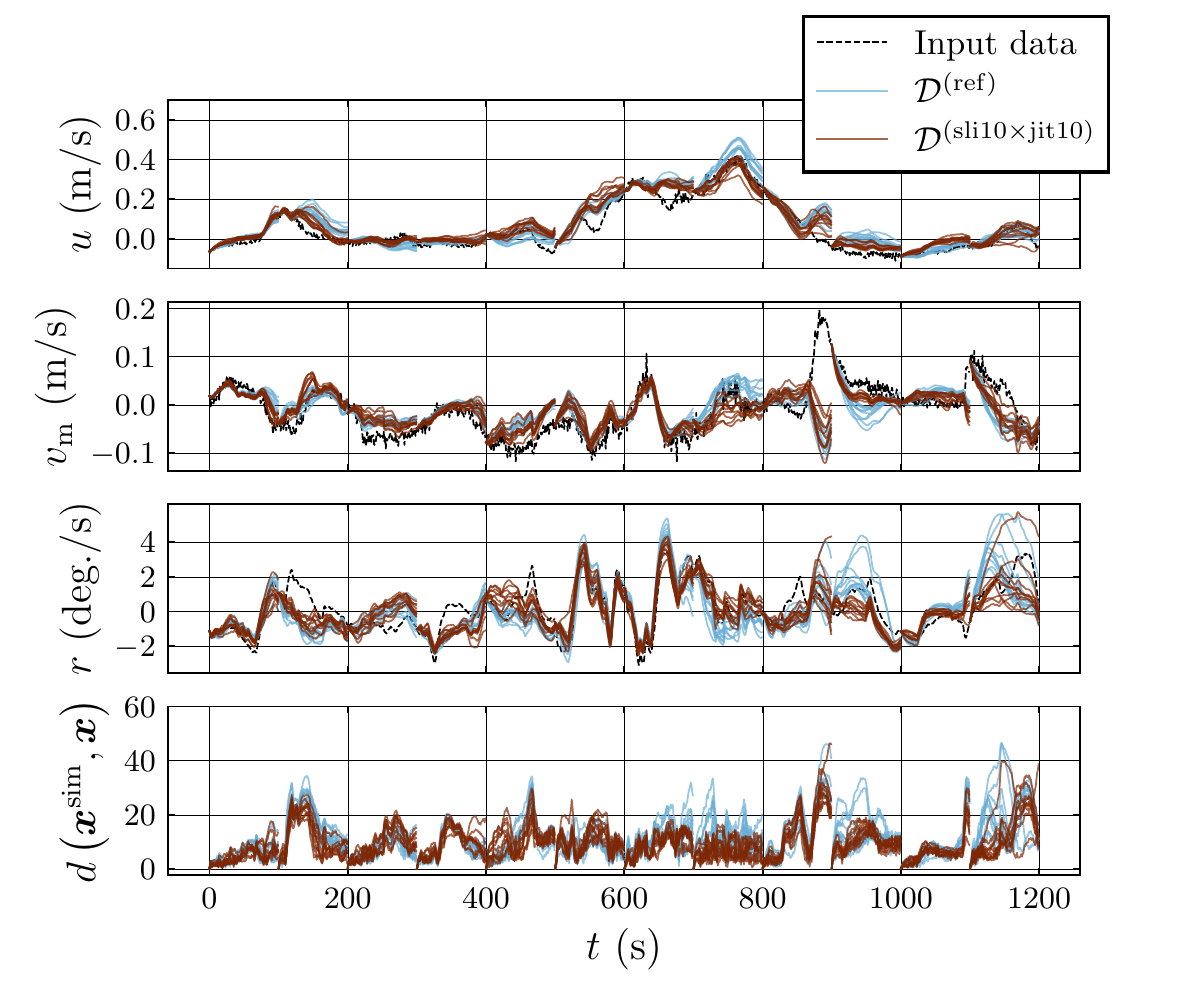}
                \caption{Prediction ship state variables $\boldsymbol{\nu}$ and error $d\left(\boldsymbol{x}^{\mathrm{sim}}, \boldsymbol{x}\right)$ using the optimal parameter trained by $\mathcal{D}^{(\text{ref})}$ and $\mathcal{D}^{(\text{sli10}\times\text{jit10})}$.}
                \label{fig:traj_base_slicing10Xjittering10_error}
            \end{figure}
        
            The prediction results of the dynamic model with optimal parameters $\boldsymbol{\theta}_{\mathrm{opt}}$ are presented here.
            The prediction error of the dynamic model on the test data was computed, and the evaluation function $\mathcal{L}\left(\boldsymbol{\theta}; \mathcal{D}^{(\text{test})} \right)$ was calculated. The obtained values of the evaluation function and its average value for random numbers are shown in \Cref{fig:test_loss}. To show the details of the prediction result, the time series of control inputs and relative wind speed and direction in the test dataset are presented in \Cref{fig:traj_supplement}, and \Cref{fig:traj_base_double_error,fig:traj_base_jittering10_error,fig:traj_base_slicing10_error,fig:traj_base_slicing10Xjittering10_error} show the time series of the ship state variables $\boldsymbol{x}$, the predicted ship state variables $\boldsymbol{x}^{\mathrm{sim}}$, and the temporary errors $d\left(\boldsymbol{x}^{\mathrm{sim}}, \boldsymbol{x}\right)$. Here, the number of time steps in the numerical simulation is $I=100$. In other words,
            the predicted ship state variables $\boldsymbol{x}^{\mathrm{sim}}$ were initialized with the measured one $\boldsymbol{x}$ every 100 time steps (100 s).

            First, we focus on the dataset augmented by slicing. \Cref{fig:test_loss} shows that the mean values of the evaluation function for $\mathcal{D}^{(\text{sli2})}$ and $\mathcal{D}^{(\text{sli10})}$ are smaller than that for $\mathcal{D}^{(\text{ref})}$. This indicates that slicing improves the generalization performance regarding $\mathcal{D}^{(\text{test})}$. However, since there is no significant difference between $\mathcal{D}^{(\text{sli10})}$ and $\mathcal{D}^{(\text{sli2})}$, increasing the amount of data augmented by slicing does not necessarily improve generalization performance.
            Next, we compare the results of $\mathcal{D}^{(\text{ref})}$ and the dataset augmented by jittering, $\mathcal{D}^{(\text{jit2})}$ or $\mathcal{D}^{(\text{jit10})}$. Although the mean values of the evaluation function for $\mathcal{D}^{(\text{jit2})}$ are only slightly smaller than those of $\mathcal{D}^{(\text{ref})}$, that for $\mathcal{D}^{(\text{jit10})}$ are even smaller. Thus, Jittering is also an augmentation method that improves the generalization performance regarding $\mathcal{D}^{(\text{test})}$.
            Then, \Cref{fig:test_loss} shows that the mean values of the evaluation function for the dataset augmented by simultaneous slicing and jittering, $\mathcal{D}^{(\text{sli2}\times\text{jit2})}$ and $\mathcal{D}^{(\text{sli10}\times\text{jit10})}$ are smaller than that for $\mathcal{D}^{(\text{ref})}$. These results indicate that slicing and jittering can be used in combination.

            However, the evaluation function values of $\mathcal{D}^{(\text{d-ref})}$ are smaller than that of any augmented dataset. Even though $\mathcal{D}^{(\text{sli10})}$ and $\mathcal{D}^{(\text{jit10})}$ have more data than $\mathcal{D}^{(\text{d-ref})}$, the prediction error trained by $\mathcal{D}^{(\text{sli10})}$ and $\mathcal{D}^{(\text{jit10})}$ is not smaller than that by $\mathcal{D}^{(\text{d-ref})}$.
            One reason for this is the difference in prediction errors occurring from 850 to 900 s. \Cref{fig:traj_base_double_error} shows that $\mathcal{D}^{(\text{d-ref})}$ reduce that prediction error, while \Cref{fig:traj_base_jittering10_error,fig:traj_base_slicing10_error,fig:traj_base_slicing10Xjittering10_error} shows that the augmented datasets cannot. \Cref{fig:traj_supplement} shows that the relatively strong apparent wind, which is around $U_{\mathrm{A}}=5.0 \ \mathrm{(m/s)}, \gamma_{\mathrm{A}}=300 \ \mathrm{(deg.)}$, occurred from 850 to 900 s. 
            \Cref{fig:wind_histgram} shows that there is a large difference between $\mathcal{D}^{(\text{ref})}$ and $\mathcal{D}^{(\text{d-ref})}$ in the amount of data around $\gamma_{\mathrm{A}}=300 \ \mathrm{(deg.)}, U_{\mathrm{A}}=5.0 \ \mathrm{(m/s)}$. Note that slicing and jittering cannot synthesize data that are not close to any data in the original dataset. Therefore, the data augmentation methods could not reduce the prediction errors occurring from 850 to 900 s. In addition, the prediction errors occurring from 1100 to 1200 s are likely caused by the same reason. As a result, the evaluation function values of $\mathcal{D}^{(\text{d-ref})}$ becomes the smallest in any other training dataset.
        
    \section{Discussion}\label{sec:discuss}
        In \Cref{subsec:result}, the results of numerical experiments to identify a dynamic model for the automatic berthing and unberthing controller were presented. In this study, the dynamic model was represented using an NN-based model, and time-series data measured in the random maneuvers were used as datasets. In numerical experiments, slicing or jittering improved the generalization performance of the dynamic model. The simultaneous use of slicing and jittering also improved the same. Therefore, they were effective data augmentation methods when the amount of measured data was limited.
        
        On the other hand, slicing and jittering cannot synthesize data that is not close to any data in the original dataset and could not improve the generalization performance of the dynamic model within the extrapolation region of the original dataset. 
        For example, strong winds that did not appear in the training data may cause a great deterioration of the generalization performance, and this deterioration cannot be avoided by slicing or jittering. Therefore, when random maneuvering tests or trials are used, it is desirable to measure the data that are widely distributed and have few extrapolation regions.
        
        However, it is impractical to observe the desired data with limited measurement time due to uncontrollable wind disturbances. To improve the generalization performance of the extrapolated state, we need to incorporate physical or hydrodynamic knowledge in addition to the information obtained from the measured data. This will be our future work.
        
    \section{Conclusion}\label{sec:conclude}
        In this study, data augmentation was introduced to the parameter identification problem to improve the generalization performance of the dynamic model for a berthing and unberthing controller. Slicing and jittering were used as data augmentation methods, and the method for applying these techniques to parameter identification problems was demonstrated. The parameters of the dynamic model were identified by minimizing the error between the measured state variables and the state variables simulated by the dynamic model. To validate the effectiveness of the data augmentation methods, numerical experiments were conducted. In numerical experiments, the dynamic model was represented by an NN-based model, and time-series data measured in free-running model tests of the random maneuvers were used as datasets. The findings of the numerical experiments are summarized as follows:
        \begin{itemize}
            \item 
            Slicing and jittering improved the generalization performance of the dynamic model when the amount of measured data was limited.
            \item 
            Slicing and jittering did not improve the generalization performance of the dynamic model within the extrapolation region of the original dataset because they cannot synthesize data that is not close to any of the measured data.
        \end{itemize}
        Therefore, we confirmed that slicing and jittering were effective data augmentation methods in those numerical experiments. 
        On the other hand, it was found necessary to collect data that are widely dispersed to reduce extrapolation regions when random maneuvering tests are used to identify a dynamic model for the automatic berthing and unberthing controller.
        
    \section*{Acknowledgements}
        This paper is a preprint published in the Journal of Marine Science and Technology, and the public version is available at (https://doi.org/10.1007/s00773-024-01036-w). 
        This study was supported by a scholarship from the Shipbuilders' Association of Japan (REDAS), and a Grant-in-Aid for Scientific Research from the Japan Society for the Promotion of Science (JSPS KAKENHI Grant \#22H01701 and \#23KJ1432).
        The authors would like to express gratitude to Japan Hamworthy \& Co., Ltd for the technical discussion, and to Enago (www.enago.jp) for reviewing the English language.
        The authors would like to thank members of the Ship Intelligentization Subarea, Osaka University, for conducting the model experiment: Hiroaki Koike, Nozomi Amano, Yuta Fueki, and Dimas M. Rachman.

    \bibliographystyle{spphys}       
    \bibliography{reference}   

\end{document}